\documentclass{PoS}

\usepackage{graphicx,amssymb,amsmath,wasysym,subfigure}
\usepackage[numbers,sort&compress]{natbib}

\title{Neutralino dark matter and naturalness of the electroweak scale}

\ShortTitle{Neutralino dark matter and naturalness of the electroweak scale}

\author{G. B\'elanger\\
        LAPTh,Universit\'e de Savoie, CNRS, 9 Chemin de Bellevue, B.P. 110, F-74941 Annecy-le-Vieux, France\\
        E-mail: \email{genevieve.belanger@lapth.cnrs.fr}}

\author{C. Delaunay\\
        LAPTh,Universit\'e de Savoie, CNRS, 9 Chemin de Bellevue, B.P. 110, F-74941 Annecy-le-Vieux, France\\
        E-mail: \email{cedric.delaunay@lapth.cnrs.fr}}

\author{\speaker{A. Goudelis}\\
        Institute of High Energy Physics, Austrian Academy of Sciences, Nikolsdorfergasse 18, 1050 Vienna, Austria\\
        E-mail: \email{andreas.goudelis@oeaw.ac.at}}

\abstract{If weak scale supersymmetry (SUSY) is to somehow explain the radiative stability of the Higgs boson mass, it is likely that non-minimal variants of SUSY models should be considered. Under the assumption that the dark matter of the universe is comprised of neutralinos, recent limits from direct detection experiments as well as dark matter abundance measurements place stringent bounds on the naturalness of  minimal supersymmetric models. We show that even non-minimal models introducing moderately decoupled new physics in order to address the Higgs boson mass problem face similar issues, with one interesting exception.}

\FullConference{18th International Conference From the Planck Scale to the Electroweak Scale \\
		 25-29 May 2015\\
		 Ioannina, Greece }

\begin{document}

%%%%%%%%%%%%%%%%%%%%%%%%%%%%%%%%%%%%%%%%%%%%%%%%%%%%%%%%%%%%%%%%%%%%%%%%%%%%%%%%%%%%%%%%%%%%%%%%%%%%%%
%%%%%%%%%%%%%%%%%%%%%%%%%%%%%%%%%%%%%%%%%%%%%%%%%%%%%%%%%%%%%%%%%%%%%%%%%%%%%%%%%%%%%%%%%%%%%%%%%%%%%%
%%%%%%%%%%%%%%%%%%%%%%%%%%%%%%%%%%%%%%%%%%%%%%%%%%%%%%%%%%%%%%%%%%%%%%%%%%%%%%%%%%%%%%%%%%%%%%%%%%%%%%
\section{Introduction}\label{sec:intro}

The existence of dark matter (DM), \textit{i.e.} of some form of non-luminous matter dominating the matter content of the universe, is by now a fairly well-grounded hypothesis in cosmology \cite{Jungman:1995df,Munoz:2003gx,Bertone:2004pz}. At present, all evidence for its existence relies on gravitational observations. Consequently, apart from some theoretical arguments such as the ``WIMP miracle'', very little is known on both the mass of the DM particles and on their coupling strength to the Standard Model (SM) ones. 

On a - seemingly - unrelated topic, the recent discovery \cite{Aad:2012tfa,Chatrchyan:2012xdj} of a $\simeq125\,$GeV Higgs boson at the LHC calls for the existence of physics beyond the SM. If the SM is to be viewed as an effective field theory (EFT) of some UV-completion, then the Higgs boson mass should lie close to the highest scale in the theory. This constitutes the notorious ``hierarchy'' problem, which requires either an unnaturally large fine-tuning of \textit{a priori} unrelated SM parameters or new dynamics to emerge close to the TeV scale in order to screen  the low-energy physics from large radiative corrections at very short distance. 

The question of dark matter and that of the naturalness of the EW scale share no obvious connection. The fact that naturalness arguments point to the existence of new physics not far above the EW scale, does by no means guarantee that this new physics should involve some particle that is electrically- and colour- neutral or stable on cosmological timescales, while DM physics itself shows, to the best of our knowledge, no preference for physics around the EW scale. 

However, in some concrete models such a connection can exist. In \cite{Belanger:2014vua}, we investigated this interplay in the Minimal Sypersymmetric Standard Model (MSSM) (\textit{cf} also \cite{Perelstein:2011tg,Perelstein:2012qg}) and its extensions by moderately decoupled new physics. In what follows we will show, in particular, that if supersymmetry is to address the EW scale naturalness issue and simultaneously provide the lightest neutralino as a viable DM candidate, then recent direct DM detection experimental results and relic density considerations essentially render the two incompatible. An interesting exception appears in extensions of the MSSM, which will be tested in the next few years.

%%%%%%%%%%%%%%%%%%%%%%%%%%%%%%%%%%%%%%%%%%%%%%%%%%%%%%%%%%%%%%%%%%%%%%%%%%%%%%%%%%%%%%%%%%%%%%%%%%%%%%
%%%%%%%%%%%%%%%%%%%%%%%%%%%%%%%%%%%%%%%%%%%%%%%%%%%%%%%%%%%%%%%%%%%%%%%%%%%%%%%%%%%%%%%%%%%%%%%%%%%%%%
%%%%%%%%%%%%%%%%%%%%%%%%%%%%%%%%%%%%%%%%%%%%%%%%%%%%%%%%%%%%%%%%%%%%%%%%%%%%%%%%%%%%%%%%%%%%%%%%%%%%%%
\section{Physics beyond the MSSM, the $\mu$ parameter and EW fine-tuning}\label{sec:bmssm}

After the discovery of the Higgs boson, and given the non-observation of stops at the LHC \cite{Chatrchyan:2013xna,Aad:2014kra}, the MSSM is known to suffer from a severe problem of fine-tuning in order to stabilize the electroweak scale under radiative corrections \cite{Hall:2011aa}. This ``little hierarchy problem'' can be alleviated in extensions of the minimal model by either $F$-terms or $D$-terms \cite{Batra:2003nj,Maloney:2004rc,Auzzi:2012dv,Bharucha:2013ela,Dimopoulos:2014aua,Ellwanger:2011mu,Ross:2012nr}, many of which invoke the existence of moderately decoupled new physics, usually around a few TeV. Given the multitude of MSSM extensions, a rather generic description of such Beyond the MSSM (BMSSM) physics affecting the Higgs sector can be given by means of effective field theory \cite{Strumia:1999jm,Casas:2003jx,Dine:2007xi}. To lowest order, this new physics can be described by a dimension 5 supersymmetric operator in the superpotential
\begin{equation}\label{eq:ops1}
\mathcal{W}_{\rm eff} = \mu H_u\cdot H_d + \frac{\lambda_1}{M}(H_u\cdot H_d)^2+\cdots\,,
\end{equation}
as well as a dimension-5 modification of the soft Lagrangian
\begin{equation}\label{eq:ops2}
\mathcal{L}^{\rm soft} = \mathcal{L}^{\rm soft}_{\rm MSSM} + \int d^2\theta\  \frac{\lambda_2}{M}X (H_u\cdot H_d)^2 + {\rm h.c.},
\end{equation}
where the ellipses in \eqref{eq:ops1} denote MSSM Yukawa interactions and operators of $\mathcal{O}(1/M^2)$ or higher, $H_{u,d}$ are the chiral superfields of the Higgs doublets, $H_u\cdot H_d=H_u^T(i\sigma_2)H_d$ denotes their antisymmetric product and $X=m_{\rm soft}\theta^2$ is a dimensionless $F$-term spurion parameterizing SUSY breaking effects.

The effective operators in equations \eqref{eq:ops1} and \eqref{eq:ops2} induce new quartic interactions in the Higgs scalar potential as well as extra Higgs-higgsino interactions. The BMSSM effects can be captured by just two parameters, which we define as
\begin{equation}
\epsilon_1 \equiv \lambda_1\mu^*/M, \ \epsilon_2 \equiv -\lambda_2m_{\rm soft}/M
\end{equation}
where $m_{\rm soft}$ is a common soft mass. We choose to work in a basis where $\mu>0$ while $M_{1,2}$ could have either sign. Moreover, we take all parameters to be real (for some interesting implications of choosing otherwise see, \textit{e.g.}, \cite{Blum:2010by}).

A first consequence of the BMSSM contribution is that the Higgs boson mass gets modified at tree-level. In particular, it is now possible to generate the Higgs mass with stops being essentially mass-degenerate with the top quark thus resolving the tension between the need for heavy stops in order to achieve the observed Higgs mass and the requirement for them to be light from naturalness arguments. Updated results on the required $\epsilon$ values in order to achieve this can be found in \cite{Belanger:2014vua}. In our analysis, we demand that the vacuum be exactly stable following the formalism of \cite{Blum:2009na}.

The $Z$ boson mass and the ratio of the Higgs vacuum expectation values, $\tan\beta$, set by the minimization conditions of the scalar potential in the vacuum, are also modified. To leading order in $\epsilon_{1,2}$, the tree-level relations read
\begin{equation}\label{eq:mZrel}
m_Z^2=
\frac{|m_{H_d}^2-m_{H_u}^2|}{\sqrt{1-\sin^22\beta}}-m_{H_u}^2-m_{H_d}^2-2\mu^2+4\epsilon_1 v^2\sin2\beta\,,
\end{equation}
and
\begin{equation}
\sin2\beta=\frac{2b}{m^2}+\frac{4v^2}{m^2}\left[\epsilon_1\left(1+4\frac{b^2}{m^4}\right)-\epsilon_2\frac{b}{m^2}\right]\,,
\end{equation}
where $m^2\equiv m_{H_u}^2+m_{H_d}^2+2\mu^2$. Below the EFT cutoff, stability of the EW scale requires all mass parameters in \eqref{eq:mZrel} to lie close to $m_Z$, unless an unnatural cancellation occurs among them. The fine-tuning associated to a parameter $p$ can be quantified via the Barbieri--Giudice measure \cite{Barbieri:1987fn}
\begin{equation}\label{eq:FTindiv}
\Delta_p \equiv \left|\frac{\partial \log m_Z^2}{\partial\log p}\right|\,.
\end{equation}
Under the assumption that all $\Delta_p$'s are independent, a global measure of fine-tuning can be obtained by summing them in quadrature
\begin{equation}\label{eq:FTglobal}
\Delta \equiv \sqrt{\Delta_{0}^2+\Delta_{\rm rad}^2}\,,\quad \Delta_0\equiv \sqrt{\sum_p \Delta_p^2}
\end{equation}
where the sum runs over $p=\mu,b,m_{H_u}^2,m_{H_d}^2,\epsilon_1,\epsilon_2$. $\Delta>1$ means an overall fine-tuning of $1/\Delta$, while $\Delta_{\rm rad}$ parameterizes the fine-tuning associated with the set of MSSM parameters which only contribute to the relation \eqref{eq:mZrel} at loop level and is typically dominated by the stop masses and mixings. If the BMSSM operators in \eqref{eq:ops1} and \eqref{eq:ops2} are exploited in order to enhance the Higgs mass and minimize $\Delta_{\rm rad}$, the overall fine-tuning is then dominated by the relative sensitivity of $m_Z^2$ to the tree-level parameters listed above. 

The analysis further reveals two important features: first, the contribution of the $\epsilon_2$ parameter is parametrically subdominant to that of $\epsilon_1$. Following this observation, as well as the fact that it is irrelevant for DM observables, we will set it to $0$ in everything that follows. Secondly, both ${\cal{O}}(\epsilon)$ effects are suppressed at large $\tan\beta$. Since higher-order corrections do generically not suffer from such a suppression, a consistent analysis to ${\cal{O}}(1/M)$ requires bounding $\tan\beta$ as
\begin{equation}\label{eq:tbmax}
\tan\beta \lesssim |\epsilon_1|^{-1}\sim \mathcal{O}(10)\,. 
\end{equation}

Before proceeding to the topic of neutralino DM in the (B)MSSM, let us stress two points that will be of interest in what follows: first, among the tree-level parameters in \eqref{eq:mZrel}, $\mu$ and $\epsilon_1$, due to their supersymmetric nature, simultaneously affect the Higgs and higgsino (and, hence, neutralino) sectors and are, thus, expected to have an impact on DM phenomenology \cite{Berg:2009mq,Bernal:2009hd,Bernal:2009jc}. Secondly, naturalness of the theory imposes a relatively low value of $\mu$, as close as possible to $m_Z$.

%%%%%%%%%%%%%%%%%%%%%%%%%%%%%%%%%%%%%%%%%%%%%%%%%%%%%%%%%%%%%%%%%%%%%%%%%%%%%%%%%%%%%%%%%%%%%%%%%%%%%%
%%%%%%%%%%%%%%%%%%%%%%%%%%%%%%%%%%%%%%%%%%%%%%%%%%%%%%%%%%%%%%%%%%%%%%%%%%%%%%%%%%%%%%%%%%%%%%%%%%%%%%
%%%%%%%%%%%%%%%%%%%%%%%%%%%%%%%%%%%%%%%%%%%%%%%%%%%%%%%%%%%%%%%%%%%%%%%%%%%%%%%%%%%%%%%%%%%%%%%%%%%%%%
\section{Neutralino dark matter in the MSSM and beyond}\label{sec:neutralino}

The most studied scenario of supersymmetric DM is that of a neutralino lightest supersymmetric particle (LSP), protected by some sufficiently well-preserved $R$-parity symmetry. Within the minimal (super-)field content, the neutralino is a linear superposition of the two neutral gauginos, $\tilde{B}$ (Bino) and $\tilde{W}^3$ (Wino), and the neutral components of the two Higgsinos, $\tilde{h}_d^0$ and $\tilde{h}_u^0$
\begin{equation}\label{eq:neutralino}
\chi_1^0 = {\cal{N}}_{11} \tilde{B} + {\cal{N}}_{12} \tilde{W}^3 + {\cal{N}}_{13} \tilde{h}_d^0 + {\cal{N}}_{14} \tilde{h}_u^0 .
\end{equation}
Its DM phenomenology is to a large extent governed by the ratio of the ${\cal{N}}_{1i}$ coefficients, which in turn depend on the hierarchy among the $\mu$ parameter and the two gaugino soft masses $M_1$ and $M_2$. For simplicity, we will focus on scenarios with a negligible Wino component, corresponding to $M_2 \gg \lbrace \mu, M_1 \rbrace$. All our conclusions remain unchanged under this assumption. 

We distinguish three limits, that of an almost pure Bino-neutralino which corresponds to the choice $M_1 \ll \mu$, that of an almost pure Higgsino ($\mu \ll M_1$) and mixed scenarios ($\mu \sim M_1$). These three scenarios exhibit radically different behaviours both from the point of view of the predicted DM abundance, but also from that of direct detection phenomenology.

Pure Binos do not pair-couple to any SM particle. As a consequence, such scenarios tend to severely overclose the universe. There are three basic ways to overcome this issue: the first is to introduce a tiny Higgsino component by slightly lowering the value of $\mu$ and to adjust the masses such that $m_{\chi_1^0} \sim m_{h/H/A}/2$, resulting in a resonance-enhanced $s$-channel annihilation cross-section (``funnel region''). The second, is to render another MSSM particle quasi mass-degenerate with the LSP and to rely on coannihilation processes (``coannihilation region''). The third is by lowering the mass of some sfermion (typically around ${\cal{O}}(100~{\rm GeV})$) and annihilating into SM fermions through $t$-channel sfermion exchange (``bulk region'').

Pure Higgsinos, on the other hand, tend to annihilate too efficiently into SM gauge bosons through $s$-channel Higgs exchange. In order to increase the predicted relic abundance in this case, we can either choose $m_{\chi_1^0} < m_W$ (corresponding to $\mu \lesssim 100$ GeV), thus kinematically forbidding annihilation into $W^+ W^-$ and $ZZ$ pairs (``light higgsino'' scenario), or $m_{\chi_1^0} \gtrsim 1$ TeV (corresponding to $\mu \sim 1$ TeV), a case in which the neutralino number density decreases to an extent that the annihilation rate can be brought down to acceptable levels (``heavy higgsino'' scenario).

Lastly, mixed Bino-Higgsino (``well-tempered'' neutralino) scenarios can indeed naturally explain the observed DM abundance in the universe \cite{ArkaniHamed:2006mb}. 

Out of these scenarios, the funnel and coannihilation regions exhibit a rather striking feature: the predicted relic density varies rapidly under small modifications of (at least from the low-energy standpoint) uncorrelated parameters of the theory. This behaviour constitutes a form of fine-tuning. The bulk region, on the other hand, relies on the existence of light sfermions which, at least for a sufficiently light Bino, is in tension with collider searches. In what follows, we will therefore decouple the sfermions from the rest of the spectrum and forbid accidental, fine-tuned relations among masses.

The spin-independent (SI) neutralino-nucleon scattering cross-section is typically dominated by SM-like Higgs boson exchange (although $H$ and $A$ can also contribute). In the case of a quasi-pure Bino, keeping terms up to $\mathcal{O}(\tan^{-2}\beta)$ and $\mathcal{O}(m_Z^2)$, the relevant coupling in the decoupling limit reads
\begin{equation}\label{eq:hXXbino}
g_{h\chi\chi}^{\tilde B-{\rm like}}\simeq \frac{2g'm_Zs_W}{\mu}\left(\frac{1}{\tan\beta}+\frac{M_1}{2\mu}-\frac{\epsilon_1 v^2}{\mu^2}\right)\,,
\end{equation}
whereas the corresponding expression for a Higgsino-like neutralino is
\begin{equation}\label{eq:hXXhino}
g_{h\chi\chi}^{\tilde h-{\rm like}}\simeq  \frac{g' m_Zs_W}{2 M_1}
\left(1+\sin2\beta-\frac{\epsilon_1v^2}{\mu^2}\cos^22\beta\right)-\sqrt{2}\frac{\epsilon_1v}{\mu}\left(1-2\sin2\beta\right)\,.
\end{equation}
From these expressions we first observe that indeed, high-purity neutralinos ($\mu \rightarrow \infty$ or $M_1 \rightarrow \infty$ respectively) tend to not couple to the Higgs boson and are, thus, expected to yield very low SI cross-sections. 

\begin{figure}[!t]
\begin{center}
\hspace{-1.2cm}
\begin{tabular}{cc} 
\includegraphics[width=0.5\textwidth]{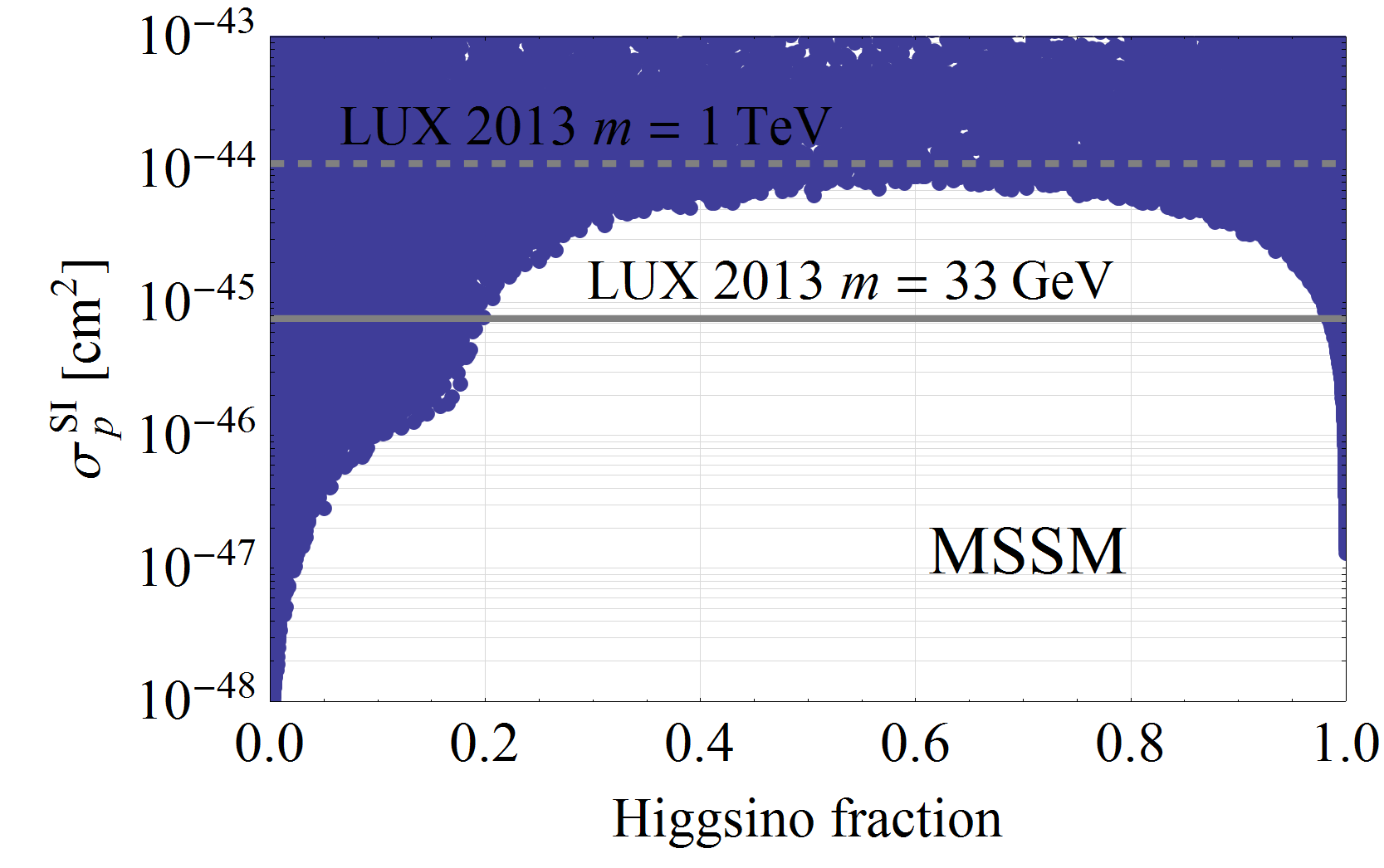}
\includegraphics[width=0.5\textwidth]{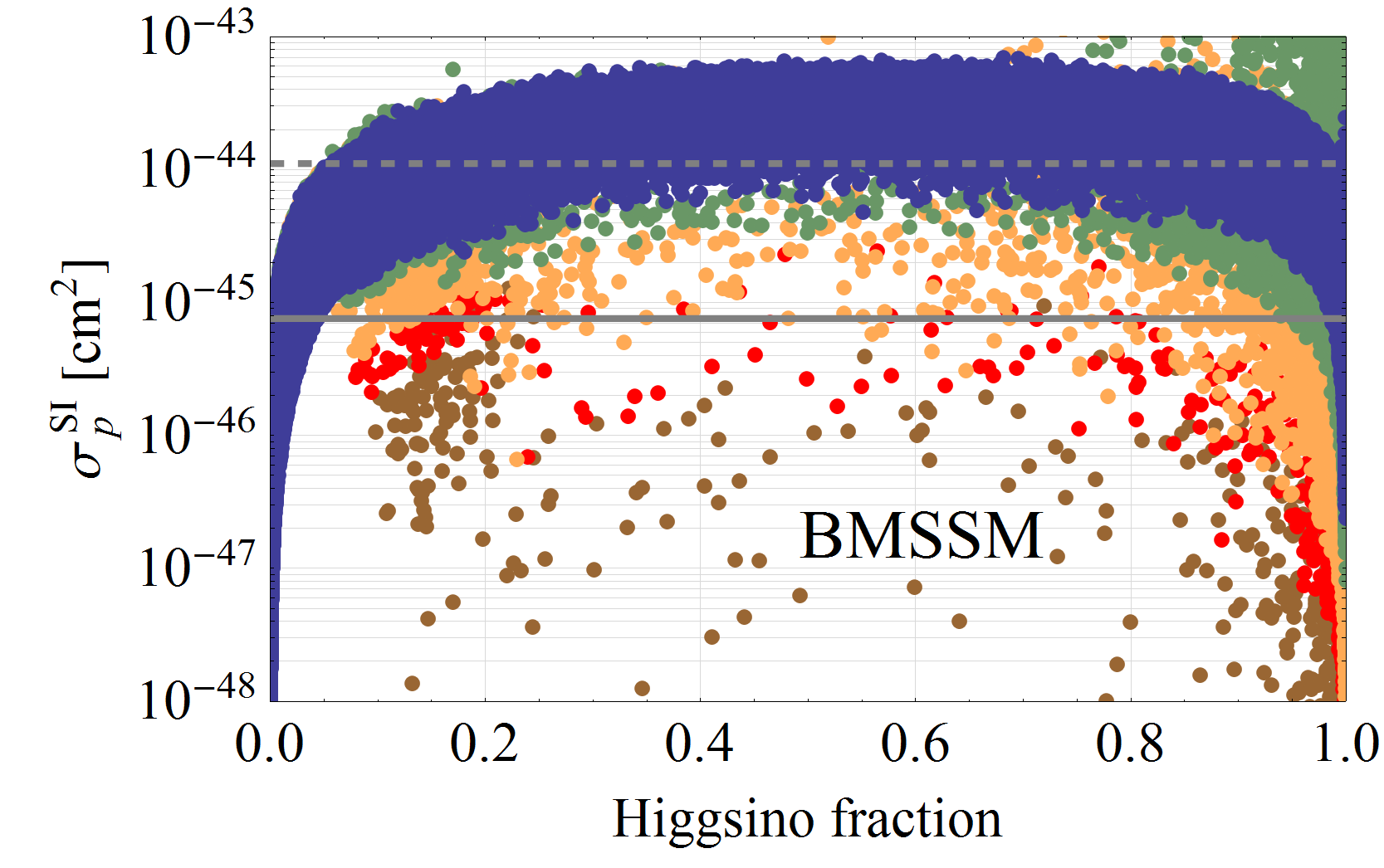}
\end{tabular}
\caption{Spin-independent cross section for DM scattering on protons as a function of the lightest neutralino Higgsino fraction in the MSSM (left panel) and the BMSSM (right panel). The solid (dashed) grey line shows the current $90\%$ CL limit from LUX \cite{Akerib:2013tjd} for $m_\chi=33\,$GeV ($1\,$TeV). In the BMSSM, colors correspond to different levels of log-sensitivity of the cross section with $\Delta_{\sigma_{\rm SI}}$ below 5 (blue), between 5 and 10 (green), 10 and 50 (orange), 50 and 100 (red) and above 100 (brown).}
\label{fig:DD_compo}
\end{center}
\end{figure}
This situation is depicted in figure \ref{fig:DD_compo} where we show the predicted SI cross-section as a function of the Higgsino fraction for the MSSM (left) and the BMSSM (right). The solid (dashed) horizontal gray lines show the latest LUX limit \cite{Akerib:2013tjd} for a WIMP mass of $30$ GeV ($1$ TeV). We see that in both models highly mixed scenarios are heavily excluded, with only extremely high-purity neutralinos being allowed. The feature appearing in the MSSM for Higgsino fractions around $0.1-0.2$ is due to the possibility for larger values of $\tan\beta$ with respect to the BMSSM, according to the limit \eqref{eq:tbmax} that we imposed on the latter. Interestingly, despite the overall similarity between the two models, in the BMSSM case we observe that there exist some mixed scenarios exhibiting very low SI cross-sections. This is due to cancellations occuring among the $u$- and $d$-quark amplitudes. However, these cancellations are purely accidental. In order to quantify the associated amount of fine-tuning, we define a measure inspired by \eqref{eq:FTindiv}
\begin{equation}\label{eq:SIFTmeasure}
\Delta \sigma_{\rm SI} \equiv \sqrt{\sum_{p} \left(\frac{d\log\sigma_{\rm SI}}{d\log p}\right)^2}\,.
\end{equation}
The coloured points in figure \ref{fig:DD_compo} correspond to fine-tuning values below 5 (blue), between 5 and 10 (green), 10 and 50 (orange), 50 and 100 (red) and above 100 (brown). 

In summary, our analysis so far shows that only quasi-pure Bino or Higgsino scenarios are compatible with direct detection constraints, implying either $\mu \gg M_1$ or $\mu \ll M_1$ respectively.

%%%%%%%%%%%%%%%%%%%%%%%%%%%%%%%%%%%%%%%%%%%%%%%%%%%%%%%%%%%%%%%%%%%%%%%%%%%%%%%%%%%%%%%%%%%%%%%%%%%%%%
%%%%%%%%%%%%%%%%%%%%%%%%%%%%%%%%%%%%%%%%%%%%%%%%%%%%%%%%%%%%%%%%%%%%%%%%%%%%%%%%%%%%%%%%%%%%%%%%%%%%%%
%%%%%%%%%%%%%%%%%%%%%%%%%%%%%%%%%%%%%%%%%%%%%%%%%%%%%%%%%%%%%%%%%%%%%%%%%%%%%%%%%%%%%%%%%%%%%%%%%%%%%%

\section{Neutralino dark matter implications for naturalness}\label{sec:results}

\begin{figure}[!t]
\begin{center}
\begin{tabular}{cc}
\includegraphics[width=0.8\textwidth]{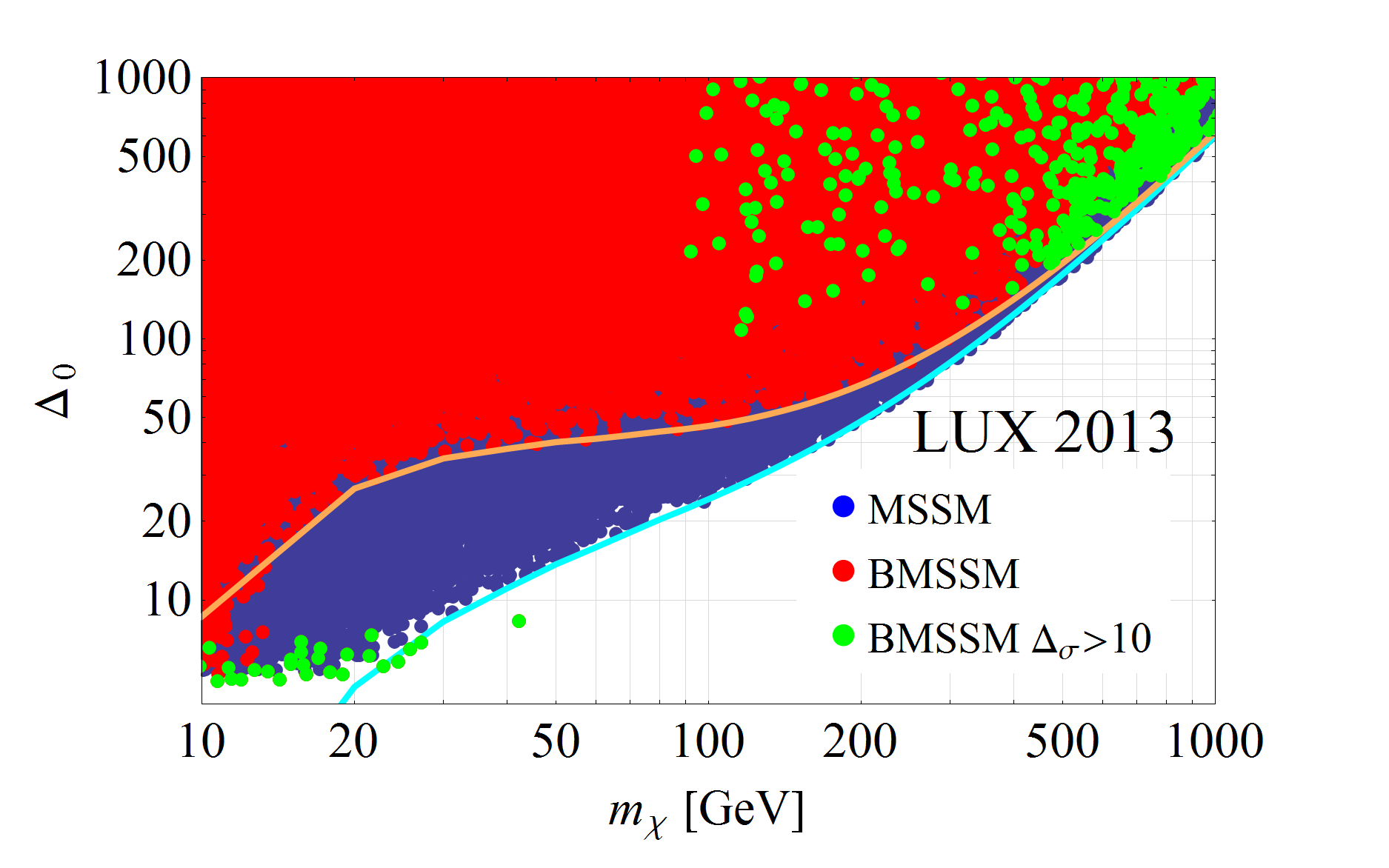}
\end{tabular}
\caption{EW fine-tuning as a function of the lightest neutralino mass for gaugino DM ($F_{\tilde H}<0.3$), imposing the current LUX limit . The low fine-tuning BMSSM points in green arise at the expense of a significant accidental cancellation in the scattering cross section of $\Delta \sigma_{\rm SI}>10$. The cyan (orange) line denotes the minimal fine-tuning in the MSSM (BMSSM).}
\label{fig:deltamchiBino}
\end{center}
\end{figure}

Direct detection constraints have important implications for the fine-tuning of neutralino DM models. Naturalness of the EW scale demands that $\mu$ be not larger than a few hundreds of GeV. For Bino scenarios this necessitates an extremely small ${\cal{O}}(10~{\rm{GeV}} )$ neutralino mass. This is demonstrated in figure \ref{fig:deltamchiBino}, where the fine-tuning is depicted as a function of the neutralino mass for the MSSM (blue points) and the BMSSM (red points).

We see that indeed, Bino-like scenarios exhibit a large amount of fine-tuning, due to the required large values of the $\mu$ parameter, unless very small masses are considered. Decreasing the value of $\mu$ in order to stabilize the EW scale would imply moving towards the mixed Bino-Higgsino regime which, as shown in figure \ref{fig:DD_compo}, is in clear conflict with the LUX results. Moreover, the BMSSM restriction \eqref{eq:tbmax} on $\tan\beta$ amounts to the fine-tuning situation being actually overall \textit{worse} than in the MSSM case. We nonetheless observe that in the BMSSM there exist some scenarios of light neutralinos (green points) exhibiting low EW fine-tuning while being compatible with direct detection results. This is due to accidental cancellations occurring in the WIMP-nucleon scattering amplitude. Quantifying the associated fine-tuning according to \eqref{eq:SIFTmeasure} reveals that these scenarios are, indeed, quite fine-tuned. Lastly, we should mention that essentially all scenarios depicted in figure \ref{fig:deltamchiBino} are incompatible with the relic density constraint, while the low mass ones are often in conflict with collider and/or Higgs invisible decay width constraints. In short, we see that in Bino-neutralino models, there is a direct contradiction between naturalness and direct detection constraints.

\begin{figure}[!t]
\begin{center}
\begin{tabular}{cc}
\includegraphics[width=0.8\textwidth]{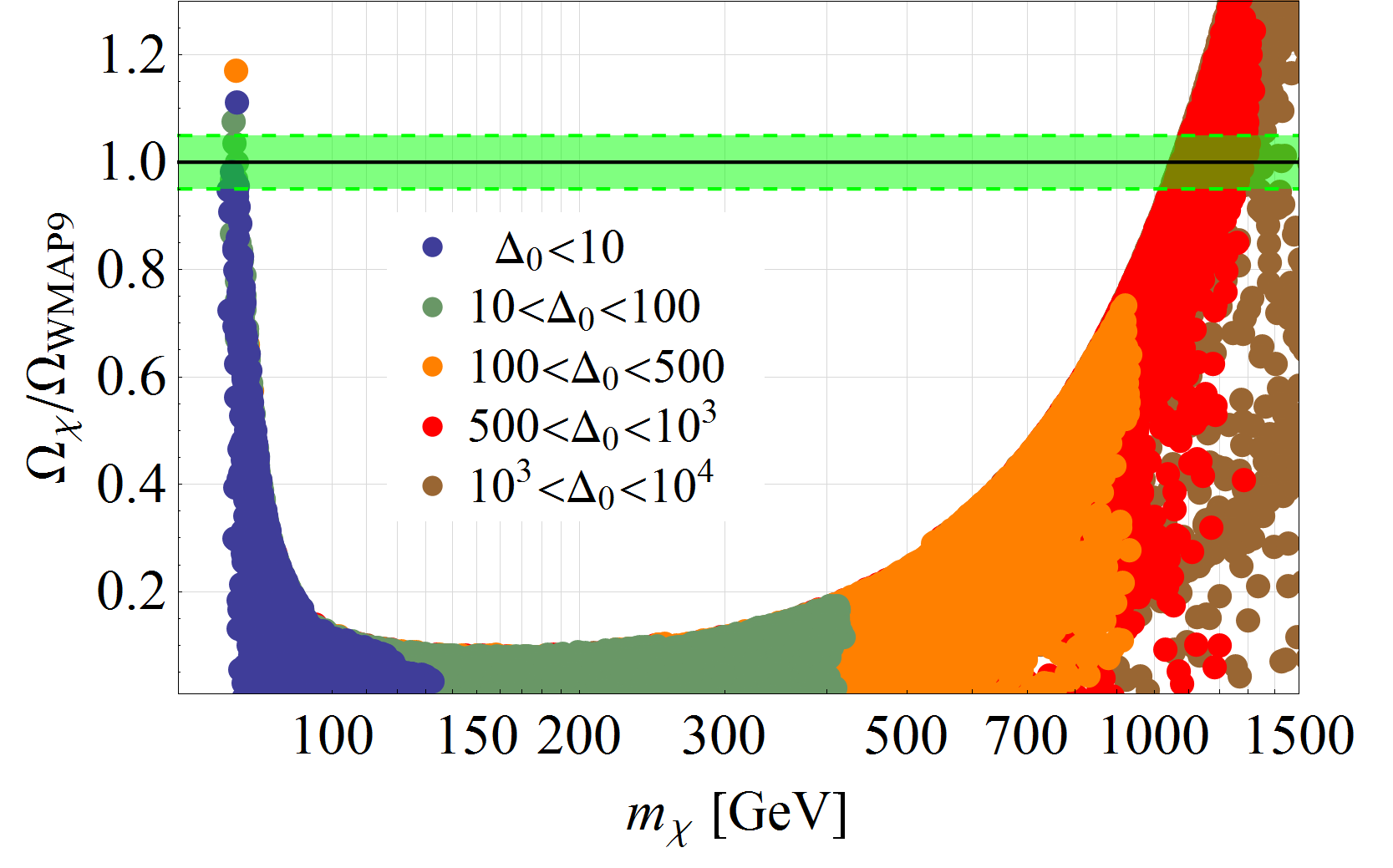}
\end{tabular}
\caption{Neutralino relic density as a function of the LSP mass for Higgsino DM. The light green band depicts the  $3\sigma$-range favored by WMAP-9 for cold DM density values. Colors denote the variation of the EW fine-tuning $\Delta_0$ with $m_\chi$. For $m_\chi\gtrsim 150\,$GeV, $\Delta_0\simeq \Delta_\mu \simeq \mathcal{O}(10)\times (m_\chi/150\,{\rm GeV})^2$.}
\label{fig:HiggsinosRelic}
\end{center}
\end{figure}

For moderate neutralino mass values, Higgsino-like scenarios are, by definition, subject to less EW fine-tuning since they correspond to $\mu \ll M_1$, with $M_1$ only substantially contributing to the overall fine-tuning once it attains values of several TeV. At the same time, according to the results presented in figure \ref{fig:DD_compo}, there is no contradiction between direct detection and naturalness in this scenario, since pure Higgsino scenarios can easily evade LUX constraints. However, these scenarios are in conflict with relic abundance considerations and in particular with the WMAP mission $9$-year results \cite{WMAP9}. This is demonstrated in figure \ref{fig:HiggsinosRelic}, where we show the ratio of the predicted/measured value of the DM density in the BMSSM, against the neutralino mass. Different colours correspond to different values of fine-tuning as described in the legend.

We can see that there are two regimes where the correct DM abundance can be attained. One characterized by $m_{\chi} < m_W$, which also corresponds to low values of $\mu$, and another one for $m_{\chi} \sim {\cal{O}}(1 {\rm TeV})$ characterized by large $\mu$ values, with the fine-tuning varying accordingly. It is clear that naturalness arguments strongly favour the low-mass regime. In the MSSM this regime is, however, excluded by LEP chargino searches. In particular, LEP-II set a bound $m_{\chi_1^\pm} \gtrsim 103$ GeV \cite{LEPcharginolimit} which could be relaxed to $98$ GeV to account for a maximal $5$ GeV uncertainty in the chargino mass matrix computation due to radiative corrections \cite{Giudice:1995np}. In the MSSM, the Higgsino-neutralino and Higgsino-chargino system is almost mass-degenerate, which implies that the low-mass Higgsino region is excluded by LEP. In the BMSSM, however, an interesting feature appears: the term \eqref{eq:ops1} contributes to the chargino mass matrix and can create, at tree-level, a mass splitting among $\chi_1^0$ and $\chi_1^\pm$ pushing the latter above the LEP-II bound.
\begin{figure}[!t]
\begin{center}
\begin{tabular}{cc}
\includegraphics[width=0.5\textwidth]{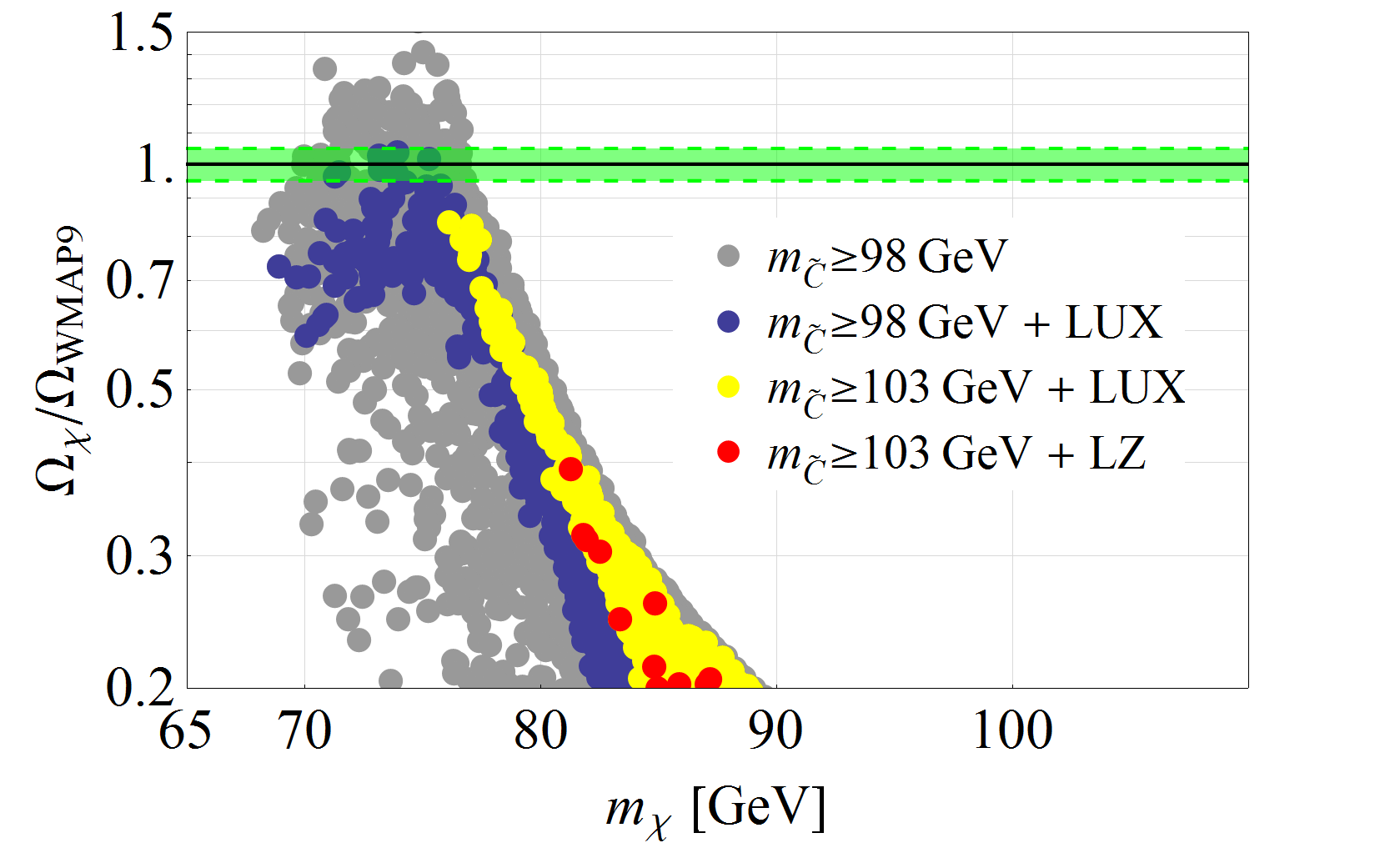}
\includegraphics[width=0.5\textwidth]{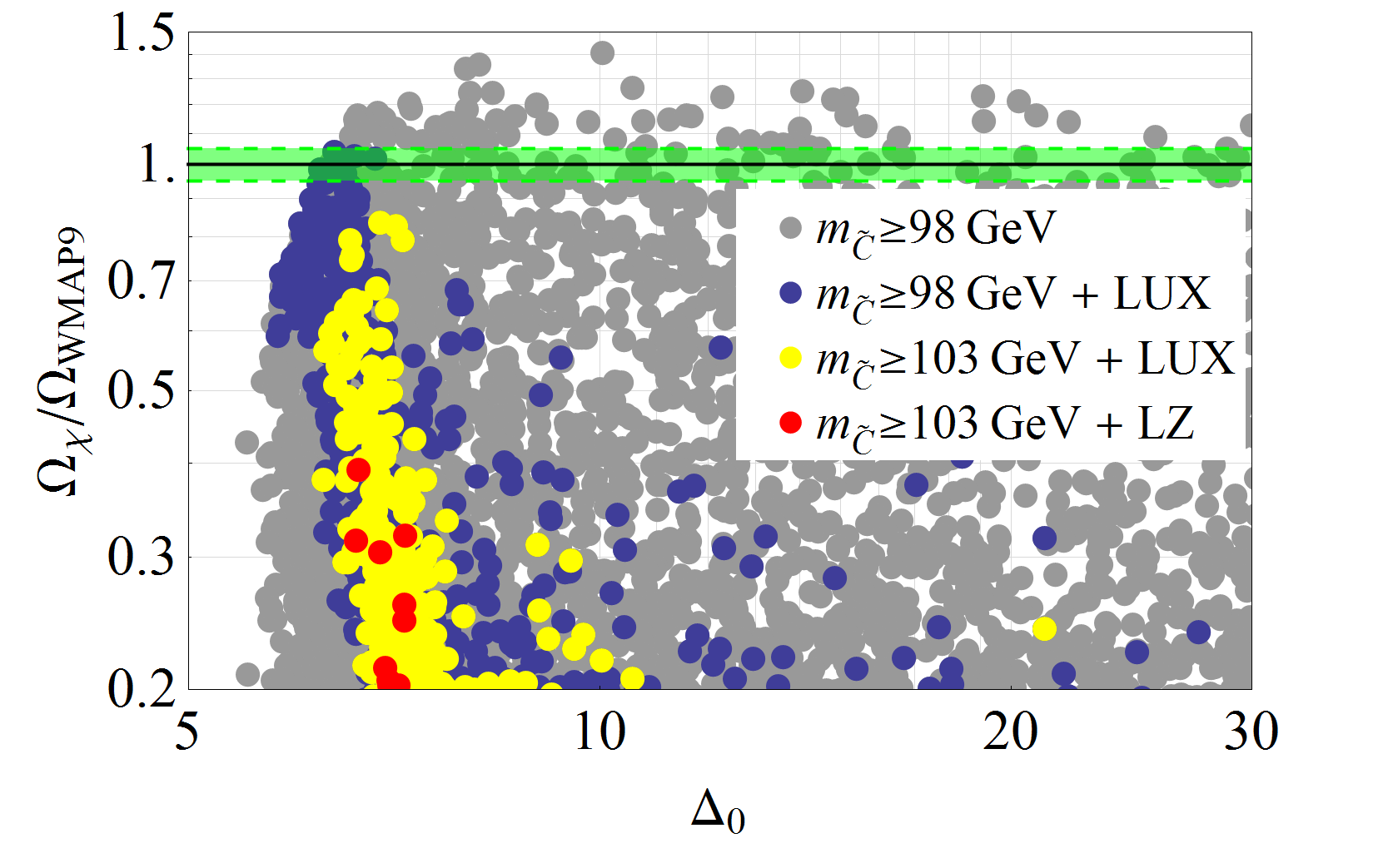}
\end{tabular}
\caption{Neutralino relic density in the BMSSM as a function of the LSP mass (left) and the EW fine-tuning $\Delta_0$ (right) for Higgsino-like LSPs below the threshold of EW boson pair production. Colors denote the requirement to satisfy various constraints on the chargino mass and the SI DM scattering cross section probed by direct searches.}
\label{fig:HiggsinosZoom}
\end{center}
\end{figure}
This is highlighted in figure \ref{fig:HiggsinosZoom}, where we zoom into the low fine-tuning region of the Higgsino-neutralino scenario, successively imposing several constraints. We indeed see that although contrived, this genuine BMSSM effect could in fact provide one of the last few possibilities for natural neutralino dark matter.
%%%%%%%%%%%%%%%%%%%%%%%%%%%%%%%%%%%%%%%%%%%%%%%%%%%%%%%%%%%%%%%%%%%%%%%%%%%%%%%%%%%%%%%%%%%%%%%%%%%%%%
%%%%%%%%%%%%%%%%%%%%%%%%%%%%%%%%%%%%%%%%%%%%%%%%%%%%%%%%%%%%%%%%%%%%%%%%%%%%%%%%%%%%%%%%%%%%%%%%%%%%%%
%%%%%%%%%%%%%%%%%%%%%%%%%%%%%%%%%%%%%%%%%%%%%%%%%%%%%%%%%%%%%%%%%%%%%%%%%%%%%%%%%%%%%%%%%%%%%%%%%%%%%%
\section{Conclusions}\label{sec:conclusions}

The fine-tuning related to the stabilization of the EW scale under radiative corrections can be addressed in extensions of the MSSM by moderately decoupled new sectors. In such scenarios, the fine-tuning associated to tree-level parameters, and in particular $\mu$, can become dominant. Due to the supersymmetric nature of the $\mu$ parameter, as well as that of the BMSSM operator in \eqref{eq:ops1}, modifications in the Higgs sector translate to modifications in the neutralino (and in particular Higgsino) sector. Much like in the MSSM, in gaugino DM scenarios EW naturalness is in conflict with direct detection constraints, since the latter push $\mu$ to values that destabilize the EW scale. In Higgsino scenarios this is not the case, but demanding for neutralinos to constitute the major part of DM in the universe according to the standard thermal freeze-out picture brings the MSSM, once again, in conflict with naturalness due to the LEP-II bounds on light charginos. This tension could be alleviated, as we showed, in extensions of the MSSM that can lift the Higgsino-neutralino and Higgsino-chargino degeneracy at tree-level. Such scenarios will be probed in the next few years, both by direct detection experiments and by LHC searches. If no signal is found, then in order for supersymmetry to address the DM problem without inducing an unacceptable level of fine-tuning of the EW scale, either the thermal history of the universe would have to be altered in a non-trivial way (by introducing, for instance, some DM regeneration mechanism in Higgsino-like scenarios with $m_\chi > m_W$) or the neutralino would have to be abandoned as a plausible DM candidate.
%%%%%%%%%%%%%%%%%%%%%%%%%%%%%%%%%%%%%%%%%%%%%%%%%%%%%%%%%%%%%%%%%%%%%%%%%%%%%%%%%%%%%%%%%%%%%%%%%%%%%%
%%%%%%%%%%%%%%%%%%%%%%%%%%%%%%%%%%%%%%%%%%%%%%%%%%%%%%%%%%%%%%%%%%%%%%%%%%%%%%%%%%%%%%%%%%%%%%%%%%%%%%
%%%%%%%%%%%%%%%%%%%%%%%%%%%%%%%%%%%%%%%%%%%%%%%%%%%%%%%%%%%%%%%%%%%%%%%%%%%%%%%%%%%%%%%%%%%%%%%%%%%%%%
\bigskip

%%%%%%%%%%%%%%%%%%%%%%%%%%%%%%%%%%%%%%%%%%%%%%%%%%%%%%%%%%%%
%\section*{Acknowledgements}
%%%%%%%%%%%%%%%%%%%%%%%%%%%%%%%%%%%%%%%%%%%%%%%%%%%%%%%%%%%%
{\bf Acknowledgements:} 
AG\ is supported by the New Frontiers program of the Austrian Academy of Sciences and would like to thank the organizers of Planck 2015 for warm hospitality during the conference


\begin{thebibliography}{99}
%\cite{Jungman:1995df}
\bibitem{Jungman:1995df}
  G.~Jungman, M.~Kamionkowski and K.~Griest,
  %``Supersymmetric dark matter,''
  Phys.\ Rept.\  {\bf 267} (1996) 195
  [hep-ph/9506380].
  %%CITATION = HEP-PH/9506380;%%
  %2634 citations counted in INSPIRE as of 30 Sep 2015
  %\cite{Munoz:2003gx}
\bibitem{Munoz:2003gx} 
  C.~Munoz,
  %``Dark matter detection in the light of recent experimental results,''
  Int.\ J.\ Mod.\ Phys.\ A {\bf 19}, 3093 (2004)
  [hep-ph/0309346].
  %%CITATION = HEP-PH/0309346;%%
  %277 citations counted in INSPIRE as of 30 Sep 2015
%\cite{Bertone:2004pz}
\bibitem{Bertone:2004pz}
  G.~Bertone, D.~Hooper and J.~Silk,
  %``Particle dark matter: Evidence, candidates and constraints,''
  Phys.\ Rept.\  {\bf 405} (2005) 279
  [hep-ph/0404175].
  %%CITATION = HEP-PH/0404175;%%
  %2125 citations counted in INSPIRE as of 30 Sep 2015
%\cite{Aad:2012tfa}
\bibitem{Aad:2012tfa}
  G.~Aad {\it et al.} [ATLAS Collaboration],
  %``Observation of a new particle in the search for the Standard Model Higgs boson with the ATLAS detector at the LHC,''
  Phys.\ Lett.\ B {\bf 716} (2012) 1
  [arXiv:1207.7214 [hep-ex]].
  %%CITATION = ARXIV:1207.7214;%%
  %4973 citations counted in INSPIRE as of 30 Sep 2015
%\cite{Chatrchyan:2012xdj}
\bibitem{Chatrchyan:2012xdj}
  S.~Chatrchyan {\it et al.} [CMS Collaboration],
  %``Observation of a new boson at a mass of 125 GeV with the CMS experiment at the LHC,''
  Phys.\ Lett.\ B {\bf 716} (2012) 30
  [arXiv:1207.7235 [hep-ex]].
  %%CITATION = ARXIV:1207.7235;%%
  %4869 citations counted in INSPIRE as of 30 Sep 2015
%\cite{Hall:2011aa}
%\cite{Belanger:2014vua}
\bibitem{Belanger:2014vua}
  G.~Belanger, C.~Delaunay and A.~Goudelis,
  %``The Dark Side of Electroweak Naturalness Beyond the MSSM,''
  JHEP {\bf 1504} (2015) 149
  [arXiv:1412.1833 [hep-ph]].
  %%CITATION = ARXIV:1412.1833;%%
%\cite{Perelstein:2011tg}
\bibitem{Perelstein:2011tg}
  M.~Perelstein and B.~Shakya,
  %``Fine-Tuning Implications of Direct Dark Matter Searches in the MSSM,''
  JHEP {\bf 1110} (2011) 142
  [arXiv:1107.5048 [hep-ph]].
  %%CITATION = ARXIV:1107.5048;%%
  %26 citations counted in INSPIRE as of 08 Oct 2015
%\cite{Perelstein:2012qg}
\bibitem{Perelstein:2012qg}
  M.~Perelstein and B.~Shakya,
  %``XENON100 implications for naturalness in the MSSM, NMSSM, and $\lambda$-supersymmetry model,''
  Phys.\ Rev.\ D {\bf 88} (2013) 7,  075003
  [arXiv:1208.0833 [hep-ph]].
  %%CITATION = ARXIV:1208.0833;%%
  %65 citations counted in INSPIRE as of 08 Oct 2015
%\cite{Chatrchyan:2013xna}
\bibitem{Chatrchyan:2013xna}
  S.~Chatrchyan {\it et al.} [CMS Collaboration],
  %``Search for top-squark pair production in the single-lepton final state in pp collisions at $\sqrt{s}$ = 8 TeV,''
  Eur.\ Phys.\ J.\ C {\bf 73} (2013) 12,  2677
  [arXiv:1308.1586 [hep-ex]].
  %%CITATION = ARXIV:1308.1586;%%
  %192 citations counted in INSPIRE as of 30 Sep 2015
%\cite{Aad:2014kra}
\bibitem{Aad:2014kra}
  G.~Aad {\it et al.} [ATLAS Collaboration],
  %``Search for top squark pair production in final states with one isolated lepton, jets, and missing transverse momentum in $\sqrt s =$8 TeV $pp$ collisions with the ATLAS detector,''
  JHEP {\bf 1411} (2014) 118
  [arXiv:1407.0583 [hep-ex]].
  %%CITATION = ARXIV:1407.0583;%%
  %113 citations counted in INSPIRE as of 30 Sep 2015
\bibitem{Hall:2011aa}
  L.~J.~Hall, D.~Pinner and J.~T.~Ruderman,
  %``A Natural SUSY Higgs Near 126 GeV,''
  JHEP {\bf 1204} (2012) 131
  [arXiv:1112.2703 [hep-ph]].
  %%CITATION = ARXIV:1112.2703;%%
  %396 citations counted in INSPIRE as of 30 Sep 2015
%\cite{Batra:2003nj}
\bibitem{Batra:2003nj}
  P.~Batra, A.~Delgado, D.~E.~Kaplan and T.~M.~P.~Tait,
  %``The Higgs mass bound in gauge extensions of the minimal supersymmetric standard model,''
  JHEP {\bf 0402} (2004) 043
  [hep-ph/0309149].
  %%CITATION = HEP-PH/0309149;%%
  %198 citations counted in INSPIRE as of 30 Sep 2015
%\cite{Maloney:2004rc}
\bibitem{Maloney:2004rc}
  A.~Maloney, A.~Pierce and J.~G.~Wacker,
  %``D-terms, unification, and the Higgs mass,''
  JHEP {\bf 0606} (2006) 034
  [hep-ph/0409127].
  %%CITATION = HEP-PH/0409127;%%
  %101 citations counted in INSPIRE as of 30 Sep 2015
%\cite{Auzzi:2012dv}
\bibitem{Auzzi:2012dv}
  R.~Auzzi, A.~Giveon, S.~B.~Gudnason and T.~Shacham,
  %``A Light Stop with Flavor in Natural SUSY,''
  JHEP {\bf 1301} (2013) 169
  [arXiv:1208.6263 [hep-ph]].
  %%CITATION = ARXIV:1208.6263;%%
  %17 citations counted in INSPIRE as of 30 Sep 2015
%\cite{Bharucha:2013ela}
\bibitem{Bharucha:2013ela}
  A.~Bharucha, A.~Goudelis and M.~McGarrie,
  %``En-gauging Naturalness,''
  Eur.\ Phys.\ J.\ C {\bf 74} (2014) 2858
  [arXiv:1310.4500 [hep-ph]].
  %%CITATION = ARXIV:1310.4500;%%
  %19 citations counted in INSPIRE as of 30 Sep 2015
%\cite{Dimopoulos:2014aua}
\bibitem{Dimopoulos:2014aua}
  S.~Dimopoulos, K.~Howe and J.~March-Russell,
  %``Maximally Natural Supersymmetry,''
  Phys.\ Rev.\ Lett.\  {\bf 113} (2014) 111802
  [arXiv:1404.7554 [hep-ph]].
  %%CITATION = ARXIV:1404.7554;%%
  %14 citations counted in INSPIRE as of 30 Sep 2015
%\cite{Ellwanger:2011mu}
\bibitem{Ellwanger:2011mu}
  U.~Ellwanger, G.~Espitalier-Noel and C.~Hugonie,
  %``Naturalness and Fine Tuning in the NMSSM: Implications of Early LHC Results,''
  JHEP {\bf 1109} (2011) 105
  [arXiv:1107.2472 [hep-ph]].
  %%CITATION = ARXIV:1107.2472;%%
  %70 citations counted in INSPIRE as of 30 Sep 2015
%\cite{Ross:2012nr}
\bibitem{Ross:2012nr}
  G.~G.~Ross, K.~Schmidt-Hoberg and F.~Staub,
  %``The Generalised NMSSM at One Loop: Fine Tuning and Phenomenology,''
  JHEP {\bf 1208} (2012) 074
  [arXiv:1205.1509 [hep-ph]].
  %%CITATION = ARXIV:1205.1509;%%
  %69 citations counted in INSPIRE as of 30 Sep 2015
  %\cite{Strumia:1999jm}
\bibitem{Strumia:1999jm}
  A.~Strumia,
  %``Bounds on Kaluza-Klein excitations of the SM vector bosons from electroweak tests,''
  Phys.\ Lett.\ B {\bf 466} (1999) 107
  [hep-ph/9906266].
  %%CITATION = HEP-PH/9906266;%%
  %166 citations counted in INSPIRE as of 30 Sep 2015
%\cite{Casas:2003jx}
\bibitem{Casas:2003jx}
  J.~A.~Casas, J.~R.~Espinosa and I.~Hidalgo,
  %``The MSSM fine tuning problem: A Way out,''
  JHEP {\bf 0401} (2004) 008
  [hep-ph/0310137].
  %%CITATION = HEP-PH/0310137;%%
  %142 citations counted in INSPIRE as of 30 Sep 2015
%\cite{Dine:2007xi}
\bibitem{Dine:2007xi}
  M.~Dine, N.~Seiberg and S.~Thomas,
  %``Higgs physics as a window beyond the MSSM (BMSSM),''
  Phys.\ Rev.\ D {\bf 76} (2007) 095004
  [arXiv:0707.0005 [hep-ph]].
  %%CITATION = ARXIV:0707.0005;%%
  %163 citations counted in INSPIRE as of 30 Sep 2015
%\cite{Blum:2010by}
\bibitem{Blum:2010by}
  K.~Blum, C.~Delaunay, M.~Losada, Y.~Nir and S.~Tulin,
  %``CP violation Beyond the MSSM: Baryogenesis and Electric Dipole Moments,''
  JHEP {\bf 1005} (2010) 101
  [arXiv:1003.2447 [hep-ph]].
  %%CITATION = ARXIV:1003.2447;%%
  %27 citations counted in INSPIRE as of 30 Sep 2015
%\cite{Blum:2009na}
\bibitem{Blum:2009na}
  K.~Blum, C.~Delaunay and Y.~Hochberg,
  %``Vacuum (Meta)Stability Beyond the MSSM,''
  Phys.\ Rev.\ D {\bf 80} (2009) 075004
  [arXiv:0905.1701 [hep-ph]].
  %%CITATION = ARXIV:0905.1701;%%
  %42 citations counted in INSPIRE as of 30 Sep 2015
%\cite{Barbieri:1987fn}
\bibitem{Barbieri:1987fn}
  R.~Barbieri and G.~F.~Giudice,
  %``Upper Bounds on Supersymmetric Particle Masses,''
  Nucl.\ Phys.\ B {\bf 306} (1988) 63.
  %%CITATION = NUPHA,B306,63;%%
  %802 citations counted in INSPIRE as of 30 Sep 2015
%\cite{Berg:2009mq}
\bibitem{Berg:2009mq}
  M.~Berg, J.~Edsjo, P.~Gondolo, E.~Lundstrom and S.~Sjors,
  %``Neutralino Dark Matter in BMSSM Effective Theory,''
  JCAP {\bf 0908} (2009) 035
  [arXiv:0906.0583 [hep-ph]].
  %%CITATION = ARXIV:0906.0583;%%
  %32 citations counted in INSPIRE as of 30 Sep 2015
%\cite{Bernal:2009hd}
\bibitem{Bernal:2009hd}
  N.~Bernal, K.~Blum, Y.~Nir and M.~Losada,
  %``BMSSM Implications for Cosmology,''
  JHEP {\bf 0908} (2009) 053
  [arXiv:0906.4696 [hep-ph]].
  %%CITATION = ARXIV:0906.4696;%%
  %25 citations counted in INSPIRE as of 30 Sep 2015
%\cite{Bernal:2009jc}
\bibitem{Bernal:2009jc}
  N.~Bernal and A.~Goudelis,
  %``Dark matter detection in the BMSSM,''
  JCAP {\bf 1003} (2010) 007
  [arXiv:0912.3905 [hep-ph]].
  %%CITATION = ARXIV:0912.3905;%%
  %33 citations counted in INSPIRE as of 30 Sep 2015
%\cite{ArkaniHamed:2006mb}
\bibitem{ArkaniHamed:2006mb}
  N.~Arkani-Hamed, A.~Delgado and G.~F.~Giudice,
  %``The Well-tempered neutralino,''
  Nucl.\ Phys.\ B {\bf 741} (2006) 108
  [hep-ph/0601041].
  %%CITATION = HEP-PH/0601041;%%
  %257 citations counted in INSPIRE as of 30 Sep 2015
%\cite{Akerib:2013tjd}
\bibitem{Akerib:2013tjd}
  D.~S.~Akerib {\it et al.} [LUX Collaboration],
  %``First results from the LUX dark matter experiment at the Sanford Underground Research Facility,''
  Phys.\ Rev.\ Lett.\  {\bf 112} (2014) 091303
  [arXiv:1310.8214 [astro-ph.CO]].
  %%CITATION = ARXIV:1310.8214;%%
  %940 citations counted in INSPIRE as of 30 Sep 2015
\bibitem{WMAP9}
  G.~Hinshaw {\it et al.} [WMAP Collaboration],
  %``Nine-year Wilkinson Microwave Anisotropy Probe (WMAP) Observations: Cosmological Parameter Results,''
  Astrophys.\ J.\ Suppl.\  {\bf 208} (2013) 19
  [arXiv:1212.5226 [astro-ph.CO]].
  %%CITATION = ARXIV:1212.5226;%%
  %940 citations counted in INSPIRE as of 30 Sep 2015
\bibitem{LEPcharginolimit}
  http://lepsusy.web.cern.ch/lepsusy/www/inoslowdmsummer02/charginolowdm\_pub.html
%\cite{Giudice:1995np}
\bibitem{Giudice:1995np}
  G.~F.~Giudice and A.~Pomarol,
  %``Mass degeneracy of the Higgsinos,''
  Phys.\ Lett.\ B {\bf 372} (1996) 253
  [hep-ph/9512337].
  %%CITATION = HEP-PH/9512337;%%
  %39 citations counted in INSPIRE as of 30 Sep 2015


  
\end{thebibliography}
\end{document}